**SWSC**

# A cosmic ray-climate link and cloud observations

Benjamin A. Laken[1,2,*], Enric Pallé[1,2], Jaša Čalogović[3], and Eimear M. Dunne[4]

[1] Instituto de Astrofisica de Canarias, Via Lactea s/n, 38205 La Laguna, Tenerife, Spain
[2] Department of Astrophysics, Faculty of Physics, Universidad de La Laguna, 38206 Tenerife, Spain
*corresponding author: e-mail: blaken@iac.es
[3] Hvar Observatory, Faculty of Geodesy, University of Zagreb, Kačićeva 26, 10000 Zagreb, Croatia
[4] Finnish Meteorological Institute, Kuopio Unit, 70100 Kuopio, Finland

## ABSTRACT

Despite over 35 years of constant satellite-based measurements of cloud, reliable evidence of a long-hypothesized link between changes in solar activity and Earth's cloud cover remains elusive. This work examines evidence of a cosmic ray cloud link from a range of sources, including satellite-based cloud measurements and long-term ground-based climatological measurements. The satellite-based studies can be divided into two categories: (1) monthly to decadal timescale analysis and (2) daily timescale epoch-superpositional (composite) analysis. The latter analyses frequently focus on sudden high-magnitude reductions in the cosmic ray flux known as Forbush Decrease events. At present, two long-term independent global satellite cloud datasets are available (ISCCP and MODIS). Although the differences between them are considerable, neither shows evidence of a solar-cloud link at either long or short timescales. Furthermore, reports of observed correlations between solar activity and cloud over the 1983–1995 period are attributed to the chance agreement between solar changes and artificially induced cloud trends. It is possible that the satellite cloud datasets and analysis methods may simply be too insensitive to detect a small solar signal. Evidence from ground-based studies suggests that some weak but statistically significant cosmic ray-cloud relationships may exist at regional scales, involving mechanisms related to the global electric circuit. However, a poor understanding of these mechanisms and their effects on cloud makes the net impacts of such links uncertain. Regardless of this, it is clear that there is no robust evidence of a widespread link between the cosmic ray flux and clouds.

## 1. Introduction

A link between solar activity and the Earth's climate was first suggested more than 200 years ago (Herschel 1801), and the idea has been repeatedly present in the scientific literature ever since (e.g., Eddy 1976). Over recent decades numerous paleo-climatological studies have revealed a wide range of statistical associations between various climate fluctuations and solar activity (Ram & Stolz 1999; Beer et al. 2000; Bond et al. 2001; Fleitmann et al. 2003; Versteegh 2005). Several hypotheses have been proposed to explain how such links may operate, but so far these remain unproven (Usoskin & Kovaltsov 2008). One of the most intriguing possibilities is that solar activity changes may indirectly modulate terrestrial cloud properties. The idea of a link between the cosmic ray (CR) flux and weather was initially proposed by Ney (1959), although it was Dickinson (1975) who first suggested a mechanism, postulating that variations in atmospheric ionization due to the solar modulation of the low-energy CR flux (<1 GeV) impinging on Earth's atmosphere may result in microphysical changes in cloud properties.

Although we focus on a hypothesized CR-cloud connection, we note that it is difficult to separate changes in the CR flux from accompanying variations in solar irradiance and the solar wind, for which numerous causal links to climate have also been proposed, including: the influence of UV spectral irradiance on stratospheric heating and dynamic stratosphere-troposphere links (Haigh 1996); UV irradiance and radiative damage to phytoplankton influencing the release of volatile precursor compounds which form sulphate aerosols over ocean environments (Kniveton et al. 2003); an amplification of total solar irradiance (TSI) variations by the addition of energy in cloud-free regions enhancing tropospheric circulation features (Meehl et al. 2008; Roy & Haigh 2010); numerous solar-related influences (including solar wind inputs) to the properties of the global electric circuit (GEC) and associated microphysical cloud changes (Tinsley 2008). Consequently, variations in the CR flux may be considered a proxy for solar activity, and thus the detection of a significant CR-climate link may not unambiguously confirm a CR-related mechanism, but could also imply a mechanism related to a covarying solar parameter.

Assuming a CR-cloud connection exists, there are various factors which could potentially account for a lack of detection of this relationship over both long and short timescales studies, including: uncertainties, artefacts and measurement limitations of the datasets; high noise levels in the data relative to the (likely low) amplitude of any solar-induced changes; the inability of studies to effectively isolate solar parameters; or the inability to isolate solar-induced changes from natural climate oscillations and periodicities.

Even without such limitations it is still possible that we may be unable to detect a clear CR-cloud relationship for several reasons. It may be the case that CR flux changes result only in small or dynamic changes to cloud over certain regions during certain cloud conditions, modifying properties such as vorticity strength rather than altering the detected cloud cover (Tinsley & Deen 1991; Tinsley et al. 2012). Alternatively, a CR-cloud relationship may be strongly second order, with variations in the CR flux simply enhancing natural cloud changes to a small extent over limited time periods when specific precursor conditions occur (Laken et al. 2010). Consequently, a detectable cloud change may not necessarily be expected to occur as a direct result of a solar activity change such as those occurring during a Forbush decrease (FD) event.

Many studies have tried to isolate the effects of CRs in the available cloud records and proxies. Pallé & Butler (2002a) made a review of the available evidence for cloud changes over decadal timescales and their possible links to solar activity. In this paper, with more than 10 years of improved observational datasets available, we intend to further examine and critique the empirical evidence for the existence of a CR flux link to cloud properties and formation processes.

## 2. Cosmic rays in the atmosphere: ionization and nucleation

Recently, results from both the SKY experiment and the CERN Cosmics Leaving Outdoor Droplets (CLOUD) experiment have confirmed that the presence of ions increases the nucleation rate of aerosols compared to binary neutral nucleation (Enghoff et al. 2011; Kirkby et al. 2011).

The SKY experiment provided proof of concept in that they found the formation rate of aerosol increased in the presence of ions generated using a 580 MeV electron beam (Enghoff et al. 2011). The CLOUD experiment investigated the nucleation process in more depth, demonstrating that the presence of a ternary vapour such as ammonia can enhance the nucleation rate more than ions. 100 ppt of ammonia led to a 100–1000-fold increase in the nucleation rate, while ground-level cosmic ray intensities were found to increase the nucleation rate by up to a factor of 10 (Kirkby et al. 2011). A high-voltage clearing field could also be used to remove all ions from the CLOUD chamber, meaning that the neutral nucleation rate could be measured and compared with nucleation enhanced by the presence of ions. A 3.5 GeV/c pion beam from the CERN Proton Synchrotron was also used to generate ionization rates of up to 80 $cm^{-3}$ $s^{-1}$, equivalent to the maximum expected values in the troposphere.

Instruments were used to measure quantities directly in CLOUD which were estimated for the SKY experiment as described in the Auxiliary Material of Enghoff et al. (2011). The CLOUD experiment measured the sulphuric acid vapour concentration using a Chemical Ion Mass Spectrometer, the ion concentration using a Gerdien counter, and the formation rate of particles at different sizes were measured using multiple instruments described in the Methods section of Kirkby et al. (2011), with 50% cutoff diameters between 1.3 nm and 12 nm. The SKY experiment used a single Condensation Particle Counter (CPC) with a 50% cutoff diameter of 4 nm to measure the concentration of aerosol within their chamber, from which the formation rate of particles at 1 nm was estimated. The ion concentration in SKY was estimated based on the energy provided by the electron beam, and the sulphuric acid concentration was estimated based on production and loss rates.

The CLOUD experiment also used an atmospheric pressure interface time-of-flight mass spectrometer (APi-TOF, Tofwerk AG) to determine the composition of the smallest charged clusters and understand which substances were participating in the nucleation process. Ammonia and dimethylamine were found to be present in these clusters, which led to the conclusion that the observed nucleation was not purely binary $H_2SO_4$-$H_2O$, and that enhancement from ternary vapours should be considered as well as enhancement from ions.

The experiments at SKY were performed at only a single temperature of 294 K and at two estimated sulphuric acid concentrations, $6 \times 10^8$ $cm^{-3}$ and $7 \times 10^8$ $cm^{-3}$. The CLOUD experiments described in Kirkby et al. (2011) spanned temperatures from 248 K to 293 K, and sulphuric acid concentrations from $7 \times 10^6$ $cm^{-3}$ to $1 \times 10^9$ $cm^{-3}$. The CLOUD experiment found that inorganic nucleation could not reproduce nucleation rates observed at boundary layer temperatures, implying that the nucleation observed by the SKY experiment may have included some organic- or amine-based ternary component.

Both experiments relate to the initial steps of particle formation, and neither group has yet published an estimate of the sensitivity of atmospheric aerosol to a change in the ion-induced nucleation rate based on laboratory measurements of ion-induced nucleation rates. Modelling studies from both Pierce & Adams (2009) and Kazil et al. (2012) concluded that global cloud condensation nuclei (CCN) would not be sensitive to changes in the ion-induced nucleation rate over a solar cycle.

It has also been suggested that the CR flux may alter cloud properties via changes in the GEC, known as the near-cloud mechanism. Far less is known about how GEC-related mechanisms may ultimately influence cloud cover (Carslaw et al. 2002). The primary source of atmospheric ionization away from terrestrial sources of radon is the CR flux and secondary particles. CR-induced ionization produces molecular cluster ions and maintains the atmosphere in a weakly conducting state, resulting in the charging of aerosol particles and cloud droplets. Consequently, a small direct current is able to flow vertically between the ionosphere (which is maintained at a potential of ~250 kV due to the global net effect of charging from thunderstorms and electrified clouds) and the Earth's surface. Due to the vertical current flow, space charge accumulates at upper and lower cloud boundaries as a result of the effective scavenging of ions within clouds by droplets, creating conductivity gradients at the cloud edges (Nicoll & Harrison 2010).

The accumulation of space charge at the boundaries of clouds may be sufficient to influence droplet-droplet collision (Khain et al. 2004), cloud droplet-particle collisions (Tinsley et al. 2000) and cloud droplet formation processes (Harrison & Ambaum 2008) thereby influencing clouds in a range of direct and indirect (dynamic) ways. It has been suggested that the inhibition of precipitation formation may enable relative increases in the amount of latent heat of freezing released, resulting in a considerable amplification of energy in storm systems (Tinsley 2010; Tinsley et al. 2012). Indeed, there is some observational evidence to suggest precipitation changes may occur following large fluctuations in the CR flux (Kniveton 2004).

### 2.1. Model predictions of the effects of the CR flux on atmospheric properties

Microphysical theories regarding CR-cloud links via ion-mediated nucleation are well developed, and several studies have attempted to incorporate these effects within atmospheric models to estimate the magnitude of potential affects to aerosols and clouds. Pierce & Adams (2009) were the first to use a general circulation model (GCM) experiment to estimate the effect of the ion-induced nucleation effect on CCN over the 11-year solar cycle. They found that CCN concentrations were insensitive to the nucleation changes induced between solar maximum and solar minimum, despite an increase in new particle formation by up to a factor of 4. They explained these results by noting that at high nucleation rates, more nanometre-sized particles are competing for condensable gases to grow to CCN sizes. Consequently, each particle grows more slowly than they would otherwise do in a situation with lower nucleation rates. Hence, under high nucleation rates the particles remain at small sizes relatively longer. At small sizes the particles have high coagulation and scavenging rates. This lowers the probability that the particles will survive to CCN sizes, and as a result it reduces the sensitivity of CCN concentrations to changes in the nucleation rate (Pierce & Adams 2009).

Snow-Kropla et al. (2011) found similar results using a global chemical transport model (CTM) with aerosol microphysics to examine the sensitivity of CCN to changes in the CR flux over the 11-year solar cycle and they also separately test the influence of FD events. Their CTM simulations estimate that the change in CCN concentrations at sizes larger than 80 nm was <0.2%. They note that for smaller particles (10 nm) the change in total number was larger, yet still always less than 1%. These results again indicate that, despite enhanced concentrations of small condensation nuclei the impact of the CR flux on clouds via an ion-induced nucleation mechanism is likely to be negligible. To estimate the potential upper limit of the possible effects of ion-induced nucleation changes to cloud properties, Dunne et al. (2012) conducted sophisticated global aerosol microphysics model experiments, whereby strong (15%) reductions in nucleation rates were simulated for 10-day periods. They found that despite these large perturbations to the model, statistically significant changes in global CCN concentrations did not occur.

The simulations discussed above concerned the connection between changes in the nucleation rate and CCN formation. However, Kazil et al. (2012) explored the ion-induced nucleation-cloud link further in a GCM, including changes in cloud properties and radiative forcing. They found that changes in atmospheric ionization during the 11-year solar cycle, and the resulting variations in aerosol formation, produced a globally asymmetric radiative forcing with a net cloud albedo effect of $-0.05$ W m$^{-2}$. When considered together, the recent results from the growing number of modelling simulations suggest that variations in ion-induced nucleation over both daily and decadal timescales are unable to significantly alter CCN or cloud properties.

# 3. Long-term correlation studies

For more than 30 years clouds have been measured from satellite platforms, providing a top-down, globally comprehensive view of cloud properties. The most notable satellite-based cloud dataset is the International Satellite Cloud Climatology Project (ISCCP) (Rossow & Schiffer 1991). This dataset is an intercalibration of irradiance measurements from a fleet of geostationary and polar orbiting weather satellites, operational since 1983. The ISCCP D1 dataset is provided at 3 h time intervals, over the entire globe, at a $2.5° \times 2.5°$ resolution. These data may provide the means for the evaluation of a link between cloud properties and solar activity. Indeed, in 1997, the first study claiming a link between the solar cycle and observed cloud changes was published by Svensmark & Friis-Christensen (1997) using a monthly version of the ISCCP dataset: this work purported to identify a positive correlation between total (1000–50 mb) ocean-area cloud cover from geostationary satellite data and the CR flux during the period 1983–1991. This observation led Svensmark & Friis-Christensen (1997) to conclude that changes in the Sun may be a more important factor contributing to decadal global temperature variations than anthropogenic emissions, immediately making the topic one of both great interest and controversy. They later expanded this point to argue that much of the global warming which occurred during the 20th century was due to low cloud changes related to cosmic rays rather than anthropogenic emissions (Svensmark 2007).

## 3.1. Examining the original satellite data evidence of the proposed CR-cloud connection

Reanalysis of the original Svensmark & Friis-Christensen (1997) results over a longer time period using a discrimination of cloud height found that the correlation was restricted to low level (>680 mb/<3.2 km) cloud only (Pallé & Butler 2000). These findings were contrary to the theoretically expected response; as solar-induced ionization changes are largest at high altitudes and latitudes, cloud changes should also be greatest at these locations if the ionization changes are the main drive of ion-induced cloud formation.

The findings of a low level restriction to the CR-cloud correlation by Pallé & Butler (2000) were later confirmed by Marsh & Svensmark (2000), hereafter referred to as MS00. A monthly time-series of globally averaged ISCCP low (>680 mb/<3.2 km) cloud and CR flux anomalies over the period of June 1983 to December 1994 similar to that presented in MS00 is shown in Figure 1a. MS00 also performed an analysis of local scale (individual ISCCP data pixel) correlations: a reproduction of these results is shown in Figure 1b. MS00 claimed that 15.8% of the globe showed a statistically significant positive correlation between low cloud changes and the CR flux, with a probability ($p$) value of achieving these results by chance of $p < .001\%$. However, we find the estimation of statistical significance ascribed to these results to be in error: MS00 based this calculation on 12-month smoothed data, from a calculation of the effective sample size (taking into account autocorrelation effects). Using the method detailed in Ripley (1987) and Neal (1993) we find the number of degrees of freedom ($df$) to be 4 in the CR flux dataset and 7 in the globally averaged low cloud dataset. Such low $df$ values are expected due to the small sample sizes and high levels of autocorrelation present in both datasets. Based on this limitation alone, the threshold significant correlation value ascribed by MS00 of $r = 0.6$ is far too low. In Figure 1c we have reanalysed the low cloud CR flux correlation of MS00 without the 12-month smoothing. These data show far lower correlation values, none of which achieve statistical significance at 4 $df$. The higher correlation values were achieved by the use of 12-month smoothing, as the short-term (<1 year) variability in the data was dampened, indicating that the higher (but still non-significant) correlations arose from the long-term variations: this is problematic to the MS00 hypothesis of a causal CR-cloud explanation for their results for reasons which will be outlined in the remainder of this section.

A large degree of artificial clustering can be seen in the correlation maps around the field of view (FOV) of geostationary satellites (Fig. 1b and c). For example, some of the strongest positive CR flux-low cloud correlations occur along the Eastern edge of the central (0° longitude) Meteosat FOV. Such structures in the data are strongly indicative of artificial errors. The existence of spurious trends in the ISCCP data relating to the geostationary satellite footprints has been reported by Campbell (2004) and Norris (2000, 2005), who also suggested that the correspondence of the MS00 correlations to these regions indicated a coincidental agreement between spurious cloud trends and the CR flux.

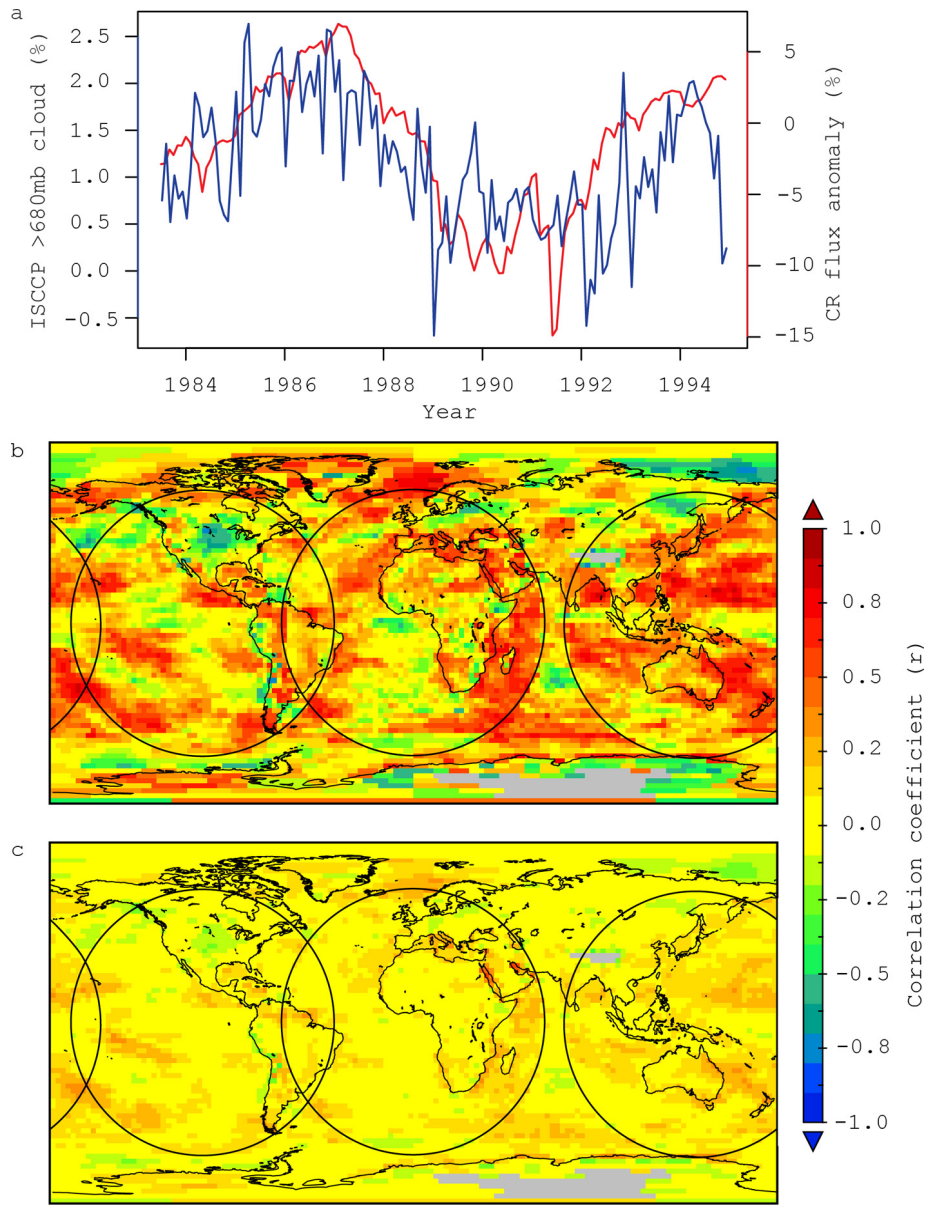

**Fig. 1.** Panel (a) shows the CR flux (red line) from combined Moscow and Climax neutron monitor data, and the globally averaged ISCCP IR low (>680 mb/<3.2 km) cloud anomaly plotted at a monthly resolution from June 1983 to December 1994, (b) shows the local correlation coefficient (*r*-values) achieved between the cloud and CR flux data for 12-month (boxcar) smoothed values. The (a–b) panels reproduce data presented by Marsh & Svensmark (2000). Panel (c) shows the *r*-values achieved from the CR flux and the ISCCP low cloud, however these values are from unsmoothed data. Panels (b–c) also indicate the positions of three geostationary satellite footprints present at the start of the ISCCP dataset. The cloud and CR data contain linear trends. From the effective sample size of the CR flux data (taking into account autocorrelation effects) it is found that there are 4 degrees of freedom (*df*) over the analysis period, under these independence constraints no pixels are found to be statistically significant ($p < .05$).

The artificial cloud anomalies were further investigated by Evan et al. (2007). These authors found that as the viewing angle of the geostationary satellite increases, the amount of estimated cloud increases in an almost linear relationship, i.e.

below the satellites the instruments have a directly overhead view of clouds, and thus can only view the cloud-tops. However, as the viewing angle increases the satellites observe cloud more obliquely, and consequently may also view the sides of the cloud and additional layers below the cloud. As a result, the instruments overestimate cloud cover towards the peripheries of their FOV.

This view-angle bias has led to an artificial long-term reduction of cloud cover in the ISCCP dataset, as over the course of the ISCCP dataset many additional geostationary satellites have been added to the network. The satellites added either replace or complement pre-existing satellite FOVs. In regions of overlapping FOVs, peripheral data is preferentially excluded, and thus cloud cover appears to decrease in these regions. Consequently, the bias towards overestimating cloud in the ISCCP dataset has become less over time (Campbell 2004; Norris 2005; Evan et al. 2007). The influence of these long-term geostationary artefacts is clearly illustrated in Figure 2, which shows the long-term trend present in the ISCCP dataset from 1983 to 2010: from this figure it is clear that ISCCP long-term cloud changes are dominated by artificial reductions in cloud. It is also important to note that the changes in satellite viewing do not occur slowly; rather, they occur as jumps in the dataset (i.e., near instantaneous changes to a new mean value). Jumps in the ISCCP data have been noted to occur in conjunction with changes such as launches of new satellites, satellite repositioning events, satellites going offline (Evan et al. 2007) and a change in the reference satellite used to perform absolute satellite intercalibrations (Knapp 2008).

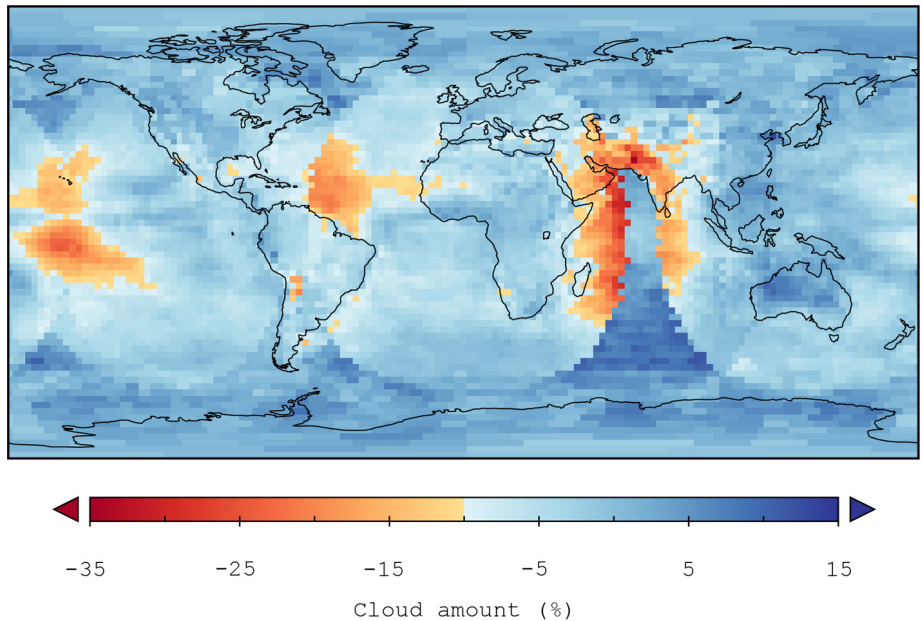

**Fig. 2.** The linear trend of each ISCCP D1 VIS-IR pixel over the 1983–2010 period. The data show clear indications of artificial trends conforming to the geostationary satellite footprint areas as noted by Campbell (2004), Norris (2000, 2005) and Evan et al. (2007).

Many reassessments of the MS00 results have been published which raise numerous further criticisms: Kristjánsson & Kristiansen (2000) and Kristjánsson et al. (2004) find that the correlation between the CR flux and low cloud cover breaks down after 1994 and does not resume, and that TSI shows a better correspondence to cloud cover changes than the CR flux; Sun & Bradley (2002) reanalyse the ISCCP cloud cover data and also examine long-term cloud data from ship observations and national weather services and find no evidence of a CR-cloud link; and Laut (2003) and Damon & Laut (2004) criticized the physical handling of the data and misleading presentation of MS00. Furthermore, although not in direct relation to the solar-cloud studies, Brest et al. (1997) state that the ISCCP data are not sensitive enough to detect small changes in cloud cover over long timescales. As the total relative uncertainties in radiance calibrations of this dataset are approximately ≤5% for visible and ≤2% for IR cloud retrievals (where absolute uncertainties are <10% and <3% respectively).

### 3.2. Low cloud data from satellites

There is a further and more fundamental issue with the satellite cloud data that makes any claim of a connection between low-cloud properties and solar activity highly suspect: we argue that it is not possible to accurately determine low cloud variations from satellite-based irradiance techniques. Our evidence for this is based on the observation, previously noted by Palle (2005), that the changes in globally averaged low cloud cover are strongly anti-correlated ($r = -0.79$) to variations in overlying cloud cover (Fig. 3). The possibility that overlying clouds may obscure low cloud, and thereby influence correlations between solar activity and cloud cover was also explored by Voiculescu et al. (2006, 2009). However, following a partial correlation analysis these authors concluded that solar-cloud correlations were not being influenced by low cloud obscurement. We disagree with this conclusion, arguing that when cloud properties are considered as a global average (Fig. 3) or over areas of frequent cloud cover (Fig. 4), the strong anti-correlation between low and middle-to-high level cloud is both clear, and statistically significant.

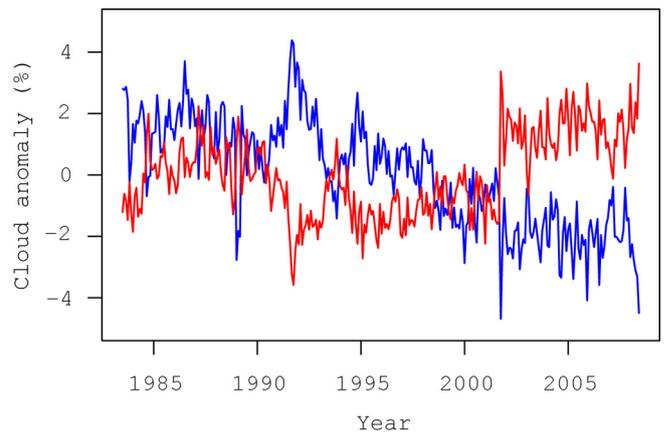

**Fig. 3.** Deseasonalized ISCCP D2 VIS-IR low (>680 mb, shown in blue) and middle + high (<680 mb, shown in red) cloud anomalies displayed over the period of 07/1983–06/2008. Data shows an anticorrelation of $r = -0.66$. Artificial jumps in the time-series are visible, most notable of which occurs after 10/2001, these appear connected to changes in the sources of data from the ISCCP satellite network.

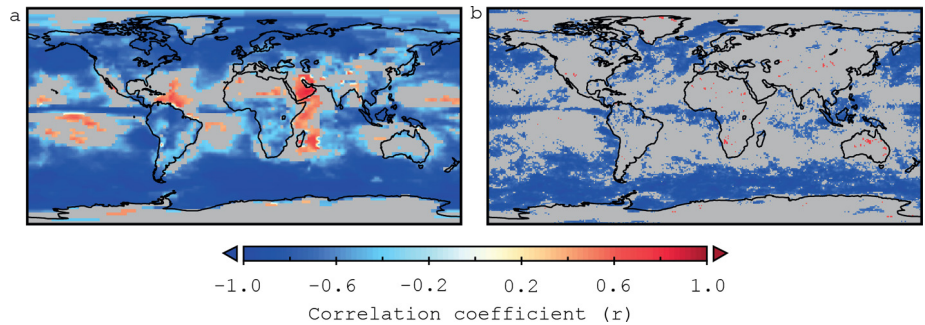

**Fig. 4.** Localized correlation coefficient (*r*-values) between middle-to-high (≤680 mb/>3.2 km) and low (>680 mb/<3.2 km) level cloud from the (a) ISCCP IR and (b) MODIS datasets. Only statistically significant *r*-values ($p < .05$) are displayed. The data shows that for regions of frequent cloud cover the low level cloud is strongly anti-correlated to overlying cloud, indicating that variations in the low cloud are predominantly artificial.

Consequently, when considering these difficulties together it is clear that we have no reliable or accurate knowledge of how global cloud cover has varied over the past several decades, and that certainly the ISCCP cloud dataset is not suitable for use in long-term correlation analysis seeking to accurately establish cloud variability and trends – a fact noted by Stordal et al. (2005). This problem is also exacerbated by climate oscillations which operate over long timescales, such as the El Niño Southern Oscillation (ENSO), which influences long-term global cloud cover and may interfere with solar-climate analysis studies (Kuang et al. 1998; Farrar 2000; Roy & Haigh 2010; Laken et al. 2012a).

### 3.3. The extended correlations: 30 years of satellite cloud data

An update to the MS00 correlation was presented by Marsh & Svensmark (2003), hereafter MS03, which showed a continued agreement between the CR flux and low cloud cover. This correspondence was achieved following the application of adjustments to the ISCCP dataset after the period of CR-cloud correlation breakdown in 1994. To achieve this effect, MS03 adjusted the ISCCP low cloud by differences they observed in the long-term trends between ISCCP and cloud estimates from the Special Sounder Microwave Instrument. While the ISCCP data may indeed require adjustments, such as those applied by Clement et al. (2009), the approach taken by MS03 is problematic, as it attempts to adjust only a portion of the ISCCP data (which does not conform to their hypothesis) against the linear trends of yet another dataset satellite-based cloud dataset. MS03 did not attempt to account for problems previously described such as view-angle bias, and long-term potentially spurious cloud trends.

The unadjusted global (90° N–90° S) ISCCP low cloud data are presented in Figure 5a, against the CR flux measured from a combination of the Moscow and Climax Colorado neutron monitors from 1983 to 2010. From Figure 5a it is clear that no good agreement exists between the low cloud data and the CR flux. A further analysis is included for the (more reliable) middle-to-high altitude cloud (≤680 mb/>3.2 km) in Figure 5b, again no agreement between the cloud data and the CR flux is evident. The correspondence between the low cloud data and the CR flux purported by MS03 is not apparent, with the updated data showing a non-significant anti-correlation of

$r = -0.15$ between monthly variations in ISCCP low cloud and CR flux. Similar conclusions were reached by Agee et al. (2012), who examined ISCCP data over the recent solar minimum (between solar cycles 23 and 24), during which time high levels of CR were recorded, and yet no corresponding cloud changes were observed to suggest a connection to solar activity.

Also plotted in Figure 5 is the low/middle-to-high level cloud as detected by the MODerate Resolution Imaging Spectroradiometer (MODIS) project (King et al. 1992). There are currently two MODIS instruments, on-board the NASA Terra and Aqua polar-orbiting satellites, operating in 36 spectral channels. These instruments measure a range of atmospheric properties and are the best globally comprehensive long-term alternative to ISSCP measurements. As evident from Figure 5 there is a considerable disagreement between the ISCCP and MODIS cloud datasets. Sources of this disagreement are attributed to differences in the way that MODIS and ISCCP define cloud, or due to the differing approaches that the datasets use to determine properties such as cloud height. Additionally, other differences are attributable to the numerous artificial errors in the ISCCP dataset as previously discussed. For a detailed comparison and evaluation of the ISCCP and MODIS cloud detection methods and cloud estimates, see Pincus et al. (2012) and references therein.

A detailed investigation of solar activity and cloud cover detected over the past decade by MODIS has been performed by Laken et al. (2012a). This work compared deseasonalized monthly MODIS cloud anomalies over the past decade to the CR flux, a UV irradiance proxy, and the TSI flux, at different cloud levels. No statistically significant correlations between solar activity and cloud cover over either global or local scales were detected. Some localized correlations between middle level (680–440 mb/3.2–6.8 km) cloud cover and the TSI flux were detected, however, these were found to be associated with the ENSO. This illustrates a major limitation of such solar-climate correlation studies: they are subject to influences from internal periodicities and fluctuations operating over similar timescales, such as ENSO, and volcanic eruptions (Farrar 2000; Roy & Haigh 2010; Laken et al. 2012a).

In addition to the ISCCP and MODIS datasets, the High-Resolution Infrared Radiometer Sounder data and Defence Satellite Meteorological Program cloud data have also been utilized to test for a solar-cloud link (e.g., Svensmark & Friis-Christensen 1997; Kuang et al. 1998). Although the temporal and spatial sampling of these datasets is more limited than the ISCCP data, it is useful to note that the datasets also do not show clear evidence of a solar-cloud link (Wylie et al. 1994; Wylie & Menzel 1999).

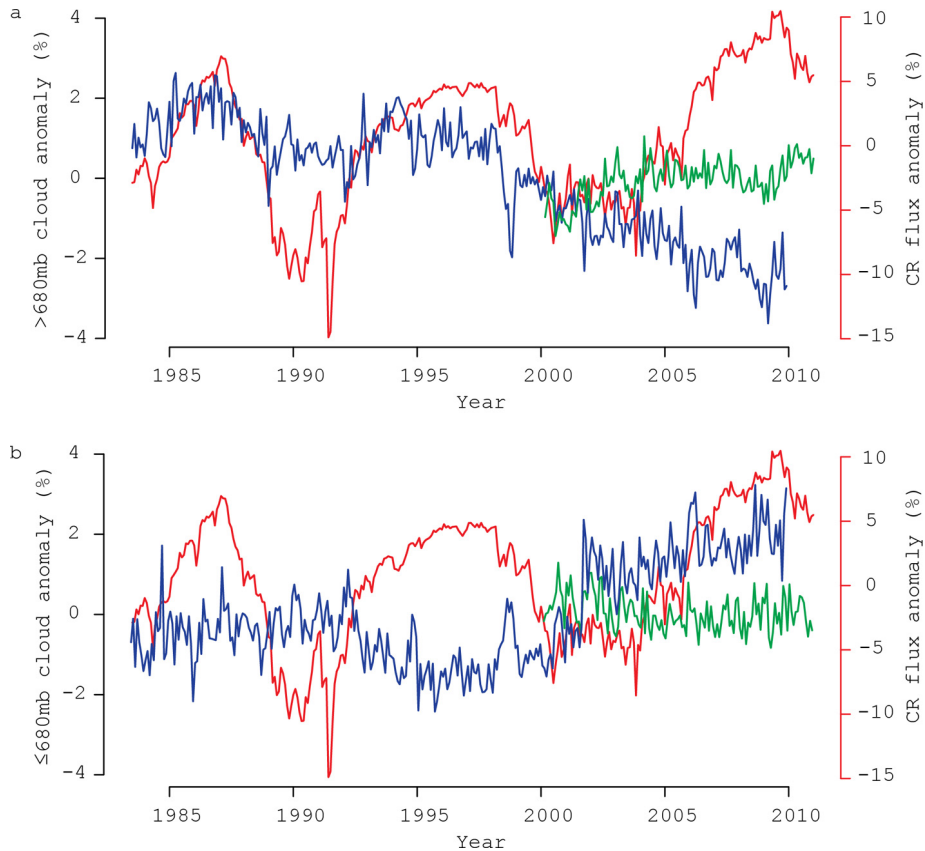

**Fig. 5.** Globally averaged cloud cover anomalies from the ISCCP IR (blue line) and MODIS (green line) cloud monitoring programmes (values on left-hand axis), operational since 1983 and 2000 respectively for: (a) low-level (>680 mb/3.2 km), and (b) middle-to-high-level (<680 mb/3.2 km) cloud cover. MODIS data is a composite of both Terra and Aqua MODIS data. Anomalies for ISCCP and MODIS are calculated against the total-period average of each dataset. Cosmic ray flux is also shown (red line, values on right-hand axis), calculated from the Climax Colorado (39.37° N, −106.18° W, 3400 m, 629 mb, 2.99 GeV) and Moscow (55.47° N, 37.32° E, 200 m, 1000 mb, 2.43 GeV) neutron monitor datasets.

## 4. FD studies

Several of the issues that plague long-term correlation studies (such as the calibration issues in the cloud data, and interference from ENSO and volcanic activity) can be overcome through the use of daily-timescale epoch-superpositional (composite) studies. Many such composite studies have been performed based around the use of high-magnitude (≥3%) sudden reductions in the CR flux termed FD events (Cane 2000). These events are produced by solar wind and interplanetary magnetic field disturbances caused by interplanetary coronal mass ejections (ICME) or corotation interaction regions (Dumbović et al. 2011). Changes in the CR flux during large FD events are of the same order of magnitude as changes experienced over the decadal solar cycle, but occur over a period of several days (Čalogović et al. 2010). The short temporal scales of FD events give these studies the benefit of being able to separate out specific changes in solar parameters. At daily timescales it is possible to distinguish changes in TSI propagating from the Sun at the speed of light, from solar-related CR flux variations that occur at the far slower speeds of propagating ICMEs. Analyses at monthly timescales do not allow this, and as a result, it is difficult to determine which (if any) of the covarying solar parameters is causally related to a solar-climate link (Kristjánsson & Kristiansen, 2000). Consequently, FD events provide an opportune means of specifically testing for a CR-cloud connection.

### 4.1. Problems with the claims that strong FD events produced a global cloud reduction

Of the FD-based studies involving satellite-era cloud datasets, only a few have reported statistically significant positive relationships. Svensmark et al. (2009, 2012) reported such a finding, claiming to identify a globally significant change in aerosol and cloud properties following FD events. These conclusions have been questioned by Laken et al. (2009) and Čalogović et al. (2010), with both studies failing to identify a statistically significant solar-cloud signal following further investigation of similar FD events. Furthermore, a detailed reanalysis showed flaws in the statistical approach of Svensmark et al. (2012) (hereafter SES12) and these flaws were also present in the Svensmark et al. (2009) study. Specifically, the statistical significance throughout their analysis was incorrectly estimated (detailed in Laken et al. 2012b). This issue arose partially from the normalization method used in their work: SES12 normalized their composites to a static averaging period preceding the FD events. This procedure is common among FD composite studies (e.g., Todd & Kniveton 2001, 2004; Dragić et al. 2011). However, the normalization procedure, when applied to raw meteorological data, creates a time-series with a non-stationary mean. As SES12 estimated their significance from data taken over the normalized period, there was an increased likelihood of obtaining false-positive results with increasing time-steps from their normalization period. This is illustrated in Figure 6 reproduced from Laken et al. (2012b) where the MODIS cloud fraction anomalies during the composite sample of SES12 are shown. The original ± 2 standard error (σ) confidence intervals calculated by SES12 are displayed on the plot (red line), and the ±2σ and ±3σ confidence intervals calculated by Laken et al. (2012b) which account for the normalization effect are displayed on the dotted and dashed lines. It is clear that during the normalization period the ±2σ confidence intervals of both groups are in close agreement. However, as the time-steps increase, the confidence intervals diverge strongly, as the SES12 ±2σ values fail to account for the increase in the spread of the mean values with increasing time from the normalization period. Consequently, the statistical significance of the observed cloud deviations following FD events reported in SES12 is highly overestimated. Similar conclusions were also drawn by Dunne (2012) who performed an independent statistical analysis of the results of SES12.

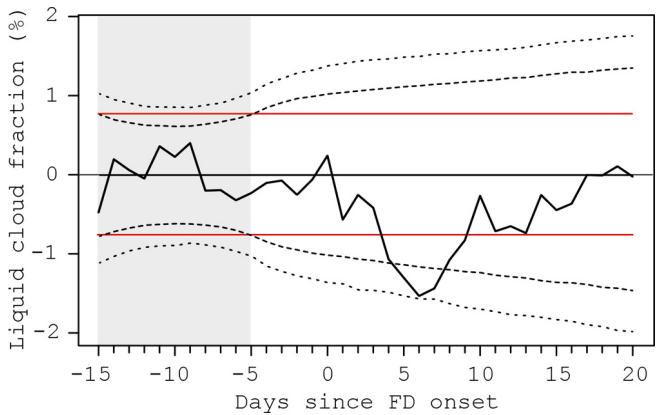

**Fig. 6.** Composite of MODIS globally averaged liquid cloud fraction anomalies (%) during the five strongest FD events of Svensmark et al. (2012). Values have been linearly de-trended, and normalized against an averaging period between days −15 and −5 (indicated by the grey shading). The ±2 standard error (σ) confidence interval calculated by Svensmark et al. (2012) is shown on the red line (calculated from 100 Monte Carlo simulations), while the ±2σ and ±3σ intervals calculated by Laken et al. (2012b) are shown on the dashed and dotted lines respectively (calculated from 10 000 Monte Carlo simulations). Figure reproduced from Laken et al. (2012b).

### 4.2. FD events and Antarctic cloud changes

Perhaps the strongest FD-based evidence from satellite-detected cloud comes from the studies of Todd & Kniveton (2001, 2004), hereafter TK. They find evidence of statistically significant cloud reductions occurring several days after the onset of FD events at stratospheric levels (180–50 mb) over the Antarctic Plateau using the ISCCP data. Due to the relative weakness of Earth's geomagnetic shielding in this region, cloud changes over such a location are tantalizing as FD events have the potential to induce maximum changes in atmospheric ionization. However, as TK note in their studies, cloud data over such locations is highly prone to detection errors. This is due to the difficulty in distinguishing a cold, bright object (i.e., a cloud) from an ice or snow covered surface: as a result of these difficulties ISCCP has been noted to mistake temperature changes for cloud changes at high latitudes (Rossow & Schiffer 1999; Laken & Pallé 2012).

In an attempt to better understand the TK results, Laken & Kniveton (2011) re-created and extended the TK composite sample, but with a crucial modification: they altered the key date of the composite from the onset of the FD event, to the date of maximal CR reduction. This is important, as the largest reductions in the CR flux can deviate from the FD onset by a period of hours to up to several days as noted by Troshichev et al. (2008). Following this change in composite methodology and the clearer isolation of the CR reduction, it was observed that cloud anomalies over the Antarctic Plateau similar to those seen by TK were actually identified two days before the maximal reductions in the CR flux.

It has been shown that compositing FD events inadvertently includes preceding reductions in TSI of ~0.4 W m$^{-2}$ (Laken et al. 2011); such TSI changes may be of significance to Earth's climate (Gray et al. 2010). The TSI reductions were found to occur two days prior to the maximal CR flux reductions. The lag time is explained by the differing transit times between irradiance changes acting at the speed of light, and solar wind disturbances (e.g., ICMEs) moving at supersonic speeds.

Based on the timing, we may question whether the TSI flux changes may be the source of the detected Antarctic Plateau cloud anomalies. A simple division of the FD composite sample into polar day and night showed that the cloud anomalies were occurring during polar darkness, and therefore it is not possible that they are causally linked to changes in TSI (Laken et al. 2011). Consequently, both the solar irradiance and the CR flux appear unconnected to these polar stratospheric cloud changes. It should also be noted that despite testing every available altitude level and grid cell of ISCCP data no other areas of statistically significant cloud anomalies were identified (Todd & Kniveton 2001, 2004; Čalogović et al. 2010; Laken & Kniveton 2011; Laken et al. 2011).

### 4.3. Searching for daily-timescale solar-cloud responses

The demonstration that FD-based composites can lead to problems in isolating CR reductions from TSI variations motivated Laken & Čalogović (2011) to construct several highly isolated daily timescale composite samples of statistically significant TSI variations which were not based on FD events. These samples unambiguously showed that cloud cover does not show any widespread anomalous variations following statistically significant changes in the TSI flux, the CR flux or a proxy for extreme UV activity within a 20-day lag period.

In the same way that the Antarctic Plateau may have been a potentially opportune location to test for a CR-cloud link due to the high variations in atmospheric ionization (resulting from weak geomagnetic shielding), certain locations have been hypothesized to be sensitive to small changes in CCN concentrations. It is in such locations where the detection of a CR-cloud relationship operating via ion-induced nucleation may be most likely (Kazil et al. 2006). This sensitivity is termed cloud susceptibility and is explained by Twomey (1991) under the aerosol indirect effect, whereby clouds in regions of low aerosol concentrations may be more prone to modification by the addition of aerosols than clouds in regions of high aerosol concentrations. A clear example of cloud susceptibility is shown by Rosenfeld et al. (2006), who demonstrated that cloud properties over aerosol-impoverished marine locations are significantly affected by the addition of relatively small amounts of aerosols from combustion in ships engines. Kristjánsson et al. (2008) adapted this idea to an investigation of FD events, performing a survey of cloud changes detected by MODIS over aerosol-impoverished oceanic regions. Overall, they found no statistically significant correlations between the CR flux and cloud properties over a composite of 22 FD events examined during an 18-day time period.

In synthesis, numerous studies exist which have not identified statistically significant relationships between satellite-detected cloud changes and FD events. While, some studies have claimed to identify a CR-cloud relationship, these findings have recently been found to be unsupported by subsequent investigation.

### 4.4. The insensitivity of FD composite methods

In FD-based composite analysis, workers are inevitably met with a choice of developing either a small sample with a select group of large events or a large sample with less event discrimination, since the number of large FD events (>8% CR flux reductions) is quite limited. The small samples have a relatively large amount of noise, but also a greater chance of detecting a signal when composite analysis methods are applied. Conversely, a large sample has a relatively smaller degree of noise, but may also have a smaller signal (due to the inclusion of weak FD events). The choice between small and large sample development was highlighted by Harrison & Ambaum (2010) in a localized ground-based study of meteorological measurements. They noted that relatively weak FD events did not appear to correspond to observable atmospheric anomalies from ground-based meteorological datasets, while high-magnitude FD-events did appear to show a meteorological response.

The majority of FD-based composite studies have small sample sizes ($n < 100$). Consequently, the signal-to-noise ratio (SNR) of such studies is likely to be low due to the high levels of noise present in meteorological datasets. As a result of this, a mechanism linking solar variations to cloud changes would need to have a high efficiency in order to be detected. However, all evidence thus far suggests that if a CR-cloud link does exist it is likely to be weak. This reasoning led Laken & Čalogović (2011) to conclude that FD-based analysis may be highly limited by issues of low SNR. This issue, coupled with the difficulties of correctly evaluating statistical significance, may have contributed to the disparities and conflicting results between various FD studies. Similar conclusions were reached by Lockwood (2012), who noted that due to the small number of FD events in composites the results are often dominated by one or two events, making the statistical significance of the composites very low. This situation was also identified as a contributor to the false significance obtained in the work of SES2 by Laken et al. (2012b), where random variations occurring during one event dominated a small composite sample.

Furthermore, there are two additional explanations which may account for why no consistent cloud cover changes may be seen with FD events: (1) a CR-cloud mechanism (ion-induced or near-cloud) is likely to be strongly second order – that is, precursor conditions are necessary before changes in the CR flux may influence cloud properties. Consequently, because FD events are random with respect to the climate system, it is not certain that cloud changes may occur following an FD event (Laken et al. 2010). (2) Alternatively, FD events might only result in dynamic effects over winter cyclogenesis regions, and therefore may not necessarily produce direct changes in cloud cover (Tinsley & Deen 1991). This may operate in the following manner: reductions in the CR flux increase atmospheric resistance to the vertical current flow of the global electric circuit, which in turn reduces the amount of space charge accumulating at cloud boundaries. Decreased space charge may result in a decrease in the electrical effect on the scavenging of CCN in clouds, and thereby influence the size distributions of cloud droplets, and processes such as coagulation and rainfall production (Tinsley 2010). As a result of changes in rainfall production, the amount of liquid water in the cloud may be modified, changing the amount of energy available for release as latent heat during freezing (Rosenfeld et al. 2008); these changes may potentially lead to significant alterations in storm vorticity strength (Tinsley et al. 2012). However, further research is needed to clearly establish the presence of this effect and its implications.

## 5. Use of ground-based cloud measurements and proxies

Certain studies have tried to overcome the difficulties of satellite-based cloud measurements through the use of ground-based datasets. These studies often benefit from data periods far longer than the satellite datasets are able to provide, giving a further means of testing the hypothesized CR-cloud connection. Types of ground-based observations include synoptic cloud extent, cloud base height observations, diurnal temperature range (DTR) variations, diffuse radiation measurements and sunshine measurements. However, these measurements are often restricted to local or regional scales, with poor and uneven geographical samplings. Some may be subject to observer biases and may be from indirect (proxy) measurements of cloud, such as sunshine measurements or DTR variations. Long-term calibrations and surface-based observing geometries may also affect ground-based observations, biasing the measurements towards low cloud.

### 5.1. Ground-based cloud observations

Ground-based synoptic cloud observations have been widely recorded over long timescales. These data are made by a human observer subjectively classifying cloud cover into a decimal or octal scale. This essentially consists of deciding how many parts in ten or eight (respectively) are covered with cloud at one or more times during the day. Although it is found that this data can include large biases at local scales (Pallé & Butler 2002b), a correctly recorded compilation of these observations may provide a relatively reliable data source (Angell et al. 1984).

In order to test a possible CR-cloud connection, Sun & Bradley (2002) made use of long-term surface-based cloud data over land from national weather services at regional scales, and over ocean from observing ships over an approximately 50 year period from the datasets of Groisman et al. (2000) and Hahn & Warren (1999). They found no correspondences to the CR flux.

Recently, a positive CR-cloud correlation was reported by Harrison et al. (2011). They detected a statistical correspondence between solar periodicities and cloud base height measured at the Lerwick observatory using a cloud base recorder. They note a reduction in the base height of low level clouds of approximately 7 m during dates of high GCR flux and based on periodogram analysis methods conclude that changes in the base height of stratiform clouds may show the presence of the 27-day and 1.68-year solar periods. Harrison et al. (2011) suggest that this signal may result from a GEC influence on cloud properties modulated by changes in the CR flux. However, although their detection is reportedly statistically significant, the data correspond to a single meteorological station: the highly localized nature of the study makes it difficult to draw broad conclusions.

### 5.2. Diffuse fraction and sunshine observations

Ground-based sunshine datasets extend over a long time period. These data provide indirect indications of cloud cover, and consequently are particularly useful for comparing to trends in the CR flux. Pallé & Butler (2001) investigated the connection between changes in records of sunshine hours recorded at sites in Ireland since the late 19th century and the 11-year solar cycle and FD events. They found no statistically significant evidence of a solar-cloud link. Pallé & Butler (2002a) extended this work, by constructing a compilation of trends from synoptic-scale regional cloud and sunshine observations over the past 150 years. They also found no evidence of corresponding long-term trends between cloud and the CR flux over the past century.

Harrison & Stephenson (2006) developed the use of these datasets further, investigating the relationship between the ratio of diffuse to total solar radiation (referred to as the diffuse fraction) to provide a proxy measurement of cloud. They compared the diffuse fraction, measured continuously at sites across the United Kingdom since 1947, to the CR flux from the Climax Colorado neutron monitor. They identified a non-linear, positive correlation between the CR flux and local cloud conditions, finding that the chance of conditions being overcast increased by approximately 19% under days of high CR flux. Their work also suggested a connection between diffuse fraction changes and FD events. This idea was further explored by Harrison & Ambaum (2010), who compared measurements of the diffuse fraction from Lerwick observatory in Shetland, UK, to FD events detected by the Climax Colorado neutron monitor. They found that for large reductions in the CR flux (of at least a 10%), there is a rapid (approximately one day) cloud response. This response was not detected for smaller FD events.

### 5.3. Diurnal temperature ranges

Using an innovative approach, Dragić et al. (2011) examined the relationship between FD events and variations in regional diurnal temperature ranges (DTR) from 189 European meteorological stations. The DTR is known to be anti-correlated to cloud cover (Dai et al. 1999) and thus DTR can be used as a proxy for cloud cover. Dragić et al. (2011) reported a statistically significant increase in the DTR of around 0.4 °C following large FD events and concluded that they had identified a statistically significant influence of the CR flux on the atmosphere. However, a re-examination of these findings by Laken et al. (2012c) showed the original results to be statistically insignificant. The over estimation of statistical significance by Dragić et al. (2011) was found to be caused by a similar normalization issue as that which affected the studies of Svensmark et al. (2009, 2012) described in section 4.1. Furthermore, based on an estimate of an upper-limit CR-cloud relationship it was found that the DTR response which Dragić et al. (2011) suggest to be causally related to CR flux is more than double the estimated upper-limit response. Consequently, it is likely that these DTR changes are attributable to stochastic noise (Laken et al. 2012c). Detailed investigation of DTR-CR flux relationships from an expanded station-based dataset, and reanalysis data found no significant associations between either the DTR and periodic (11-year and 1.68-year) solar variations, or FD events at global or regional scales (Laken et al. 2012c).

## 6. Conclusions

In this paper we have examined the evidence of a CR-cloud relationship from direct and indirect observations of cloud recorded from satellite- and ground-based measurement techniques. Overall, the current satellite cloud datasets do not provide evidence supporting the existence of a solar-cloud link. Although some positive evidence exists in ground-based studies, these are all from highly localized data and are suggested to operate via global electric circuit based mechanisms: the effects of which may depend on numerous factors and vary greatly from one location to the next. Consequently, it is unclear what the overall implications of these localized findings are. By virtue of a lack of strong evidence detected from the numerous satellite- and ground-based studies, it is clear that if a solar-cloud link exists the effects are likely to be low amplitude and could not have contributed appreciably to recent anthropogenic climate changes.

*Acknowledgements.* The authors thank Jeffrey Pierce (Dalhousie University), Brian Tinsley (University of Texas at Dallas) and two anonymous reviewers for comments. The ISCCP D1 data are available from the ISCCP website at http://isccp.giss.nasa.gov/, maintained by the ISCCP research group at the NASA Goddard Institute for Space Studies. The MODIS data were obtained from the NASA website http://ladsweb.nascom.nasa.gov. Cosmic ray data were obtained from the Solar Terrestrial physics division of IZMIRAN from http://helios.izmiran.rssi.ru. MG II core-to-wing ratio data were obtained from NOAA's National Weather Service Space Weather Prediction Center (www.swpc.noaa.gov). The author acknowledges the PMOD dataset (version d41_62_1102): PMOD/ WRC, Davos, Switzerland), which also comprises unpublished data from the VIRGO experiment on the ESA/NASA mission SoHO. The authors acknowledge the European COST Action ES1005 and support from the Spanish MICIIN, Grant No. #CGL2009-10641.